\documentclass[pra,letterpaper,onecolumn,superscriptaddress,floatfix]{revtex4}

\usepackage{graphicx,psfrag,amsmath,amssymb,amsfonts,bbm,latexsym,color,dcolumn,bm}

\begin{document}

\title{Simulating sympathetic cooling of atomic mixtures in nonlinear traps}

\author{Francisco Jauffred}

\affiliation{\mbox{Department of Physics, University of Massachusetts, Boston, MA 02125, USA}}

\author{Roberto Onofrio}

\affiliation{\mbox{Dipartimento di Fisica e Astronomia ``Galileo Galilei'', Universit\`a  di Padova, 
Via Marzolo 8, Padova 35131, Italy}}

\affiliation{\mbox{Department of Physics and Astronomy, Dartmouth College, 6127 Wilder Laboratory, 
Hanover, NH 03755, USA}}

\author{Bala Sundaram}

\affiliation{\mbox{Department of Physics, University of Massachusetts, Boston, MA 02125, USA}}

\begin{abstract}
We discuss the dynamics of sympathetic cooling of atomic mixtures in realistic, nonlinear trapping 
potentials using a microscopic effective model developed earlier for harmonic traps. 
We contrast the effectiveness of different atomic traps, such as Ioffe-Pritchard 
magnetic traps and optical dipole traps, and the role their intrinsic nonlinearity plays 
in speeding up or slowing down thermalization between the two atomic species. This discussion
includes cases of configurations with lower effective dimensionality. From a more theoretical 
standpoint, our results provide the first exploration of a generalized Caldeira-Leggett model with 
nonlinearities both in the trapping potential as well as in the interspecies interactions, 
and no limitations on their coupling strength. 
\end{abstract}

\maketitle

\section{Introduction}

Recent progress in cooling fermions to the deep quantum degenerate regime has allowed the 
exploration of unique physics such as the crossover from the Bardeen-Cooper-Schrieffer  (BCS) superfluid regime of 
Cooper pairs to the Bose-Einstein condensate (BEC) regime of tightly confined fermionic pairs~\cite{Zwierlein,Ketterlerev}. 
This result is a major breakthrough in atomic physics with many interdisciplinary implications, from high-T${}_c$ 
superconductivity to neutron stars~\cite{Chen,Giorgini,Ayryan}. 
The continued exploration of even deeper Fermi regimes  will expose quantum phase transitions and superconducting 
phases beyond the standard BCS scenario~\cite{McKayrev,Onofriorev}. It is therefore not surprising that 
a considerable number of recent experiments have focused on trapping and cooling ever larger number of fermions through 
a variety of techniques~\cite{Duarte,McKay,Burchianti,Gross}. The most successful and widespread technique is
the sympathetic cooling of fermions by using a large Bose gas reservoir. However, the realization of novel behaviour alluded
to earlier depends on the ability to optimize sympathetic cooling of Bose-Fermi mixtures.  This effort includes attempts 
to maximize the number of available bosons~\cite{Chaudhuri,Shibata}, where a crucial stage is the transfer from
a magneto-optical trap into the conservative potential provided by either a magnetic trap or an optical dipole trap. 
Currently, loading efficiencies are around 10$~\%$ providing considerable room for improvement. 
Increased numbers leading to larger bosonic clouds will also be more efficient in driving 
the fermionic counterpart into a quantum degenerate regime. Besides these practical considerations, 
Bose-Fermi mixtures are important prototypical systems, in which fundamental questions of 
thermalization, mixing, phase separation of general interest, especially in chemical physics, 
can be addressed in a pristine and controllable environment~\cite{Ufrecht}.

Our earlier work~\cite{OnoSun} introduced an effective microscopic model to discuss thermalization dynamics 
between two atomic species in the nondegenerate regime. A key ingredient was an interaction term intended to
extend Caldeira-Leggett models~\cite{Magalinskii,Ullersma1,Ullersma2,Ullersma3,Ullersma4,Caldeira1,Caldeira2,Caldeira3} 
to traps confining a finite number of atoms and arbitrary inter-species interaction strength. 
The model was later generalized to the more realistic and richer three-dimensional setting~\cite{JauOnoSun}. 
In both instances, the trapping potential used was the idealized harmonic one, though a number of nontrivial 
effects resulted from the nonlinear character of the interatomic interaction. 
These included scaling with respect to the number of atoms, remiscent of Kolmogorov scaling in fluid mixing, and
a mode locking mechanism which occurs in the presence of number-asymmetry in the two species.
However, harmonic trapping does not realistically describe the early stage of atomic transfer, more 
manifestly because it assumes an infinite energy depth. The absence of a finite energy scale and the fixed 
oscillation period ensures that the thermalization dynamics is invariant with respect to an arbitrary 
common rescaling of the temperatures of the two species. Realistic trapping potentials have instead finite 
trapping depths and consequent breaking of the energy scaling invariance, which implies that one needs to specify
the thermalization dynamics for each set of initial temperatures of the gases. 

In this paper, we focus on nonlinear traps, with special emphasis on Ioffe-Pritchard 
magnetic traps and optical dipole traps, and their impact on the dynamics of thermalization. 
For this study, realistic insights require dealing with the full three-dimensional case, as 
one-dimensional analyses are too restrictive especially in the case of a Ioffe-Pritchard trap. 
Effective lower dimensional dynamics can be realized by using parameters corresponding to anisotropic traps.
The interplay between nonlinearities of the trapping potential and the interaction term, responsible for thermalization, 
is rather subtle and, while we illustrate this with concrete examples, we do not consider the analysis to be exhaustive. 
Anharmonicities in the trapping potential usually weaken the local trapping strength but also provide channels 
of mixing and ergodicity since the oscillation periods depend on the energy of the particles. 
Which feature will prevail depends on the concrete experimental setting and dedicated 
simulations should be considered for each specific case. 

\section{Nonlinear Traps}

In order to gain quantitative insights, we focus on a magnetic trap (MT) of the Ioffe-Pritchard type, due to 
its simplicity and widespread use. In this case the magnetic field vector is expressed as~\cite{Bergeman,Ketterle}

\begin{equation}
{\bf B}=B_0 \left(\begin{array} {c}  0  \\  0  \\ 1 \end{array}\right)
+          B'  \left(\begin{array} {c}  x  \\ -y  \\ 0 \end{array}\right)
+\frac{B''}{2} \left(\begin{array} {c} -xz \\ -yz \\ z^2-\frac{1}{2}(x^2+y^2) \end{array}\right),
\tag{1}
\label{MagField}
\end{equation}
which corresponds to an amplitude
\begin{equation}
B(x,y,z)=\left[
\left(B'-\frac{B''}{2}z\right)^2x^2+\left(B'+\frac{B''}{2}z\right)^2y^2+
\left(B_0+\frac{B''}{2}(z^2-\frac{1}{2}(x^2+y^2))\right)^2
\right]^{1/2},
\tag{2}
\end{equation}
where the associated potential energy $V_{\rm mag}=- {\mathbf \mu} \cdot {\bf B}$. 
The magnetic dipole force acts on the trappable atoms (with magnetic moment antiparallel to the magnetic field) 
along the three Cartesian components. The components of the force acting on a generic magnetic moment in a 
Ioffe-Pritchard trap can be written as 

\begin{equation}
F_x=-\mu \frac{\left(B'-\frac{B''z}{2}\right)^2-{\frac{1}{2}}\left[B_0+\frac{B''}{2}
[z^2-\frac{1}{2}(x^2+y^2)]\right]B''}
{\left[\left(B'-\frac{B''z}{2}\right)^2x^2+\left(B'+\frac{B''z}{2}\right)^2y^2+
\left[B_0+\frac{B''}{2}[z^2-\frac{1}{2}(x^2+y^2)]\right]^2\right]^{1/2}}x,
\tag{3a}
\label{3a}
\end{equation}

\begin{equation}
F_y=-\mu \frac{\left(B'+\frac{B''z}{2}\right)^2-{\frac{1}{2}}\left[B_0+\frac{B''}{2}
[z^2-\frac{1}{2}(x^2+y^2)]\right]B''}
{\left[\left(B'-\frac{B''z}{2}\right)^2x^2+\left(B'+\frac{B''z}{2}\right)^2y^2+
\left[B_0+\frac{B''}{2}[z^2-\frac{1}{2}(x^2+y^2)]\right]^2\right]^{1/2}}y,
\tag{3b}
\label{3b}
\end{equation}

\begin{equation}
F_z=-\mu B'' \frac{-\frac{1}{2}\left(B'-\frac{B''z}{2}\right)x^2+
\frac{1}{2}\left(B'+\frac{B''z}{2}\right)y^2
+\left[B_0+\frac{B''}{2}[z^2-\frac{1}{2}(x^2+y^2)]\right]z}
{\left[\left(B'-\frac{B''z}{2}\right)^2x^2+\left(B'+\frac{B''z}{2}\right)^2y^2+
\left[B_0+\frac{B''}{2}[z^2-\frac{1}{2}(x^2+y^2)]\right]^2\right]^{1/2}}.
\tag{3c}
\label{3c}
\end{equation} 

\begin{figure}[b]
\includegraphics[width=0.46\textwidth, clip=true]{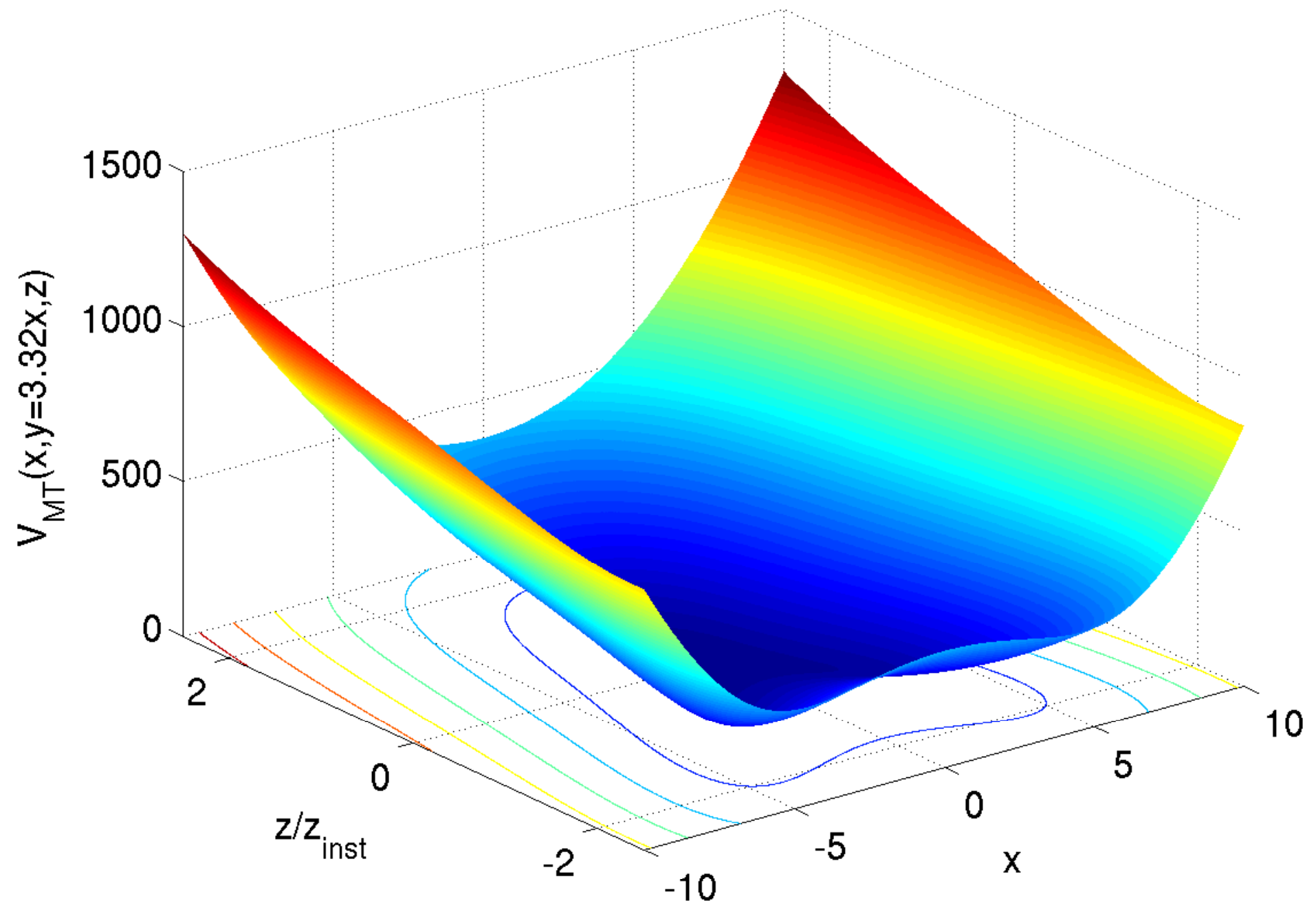}
\caption{Illustration of spatially asymmetric features of the magnetic trap potential. 
An arbitrary slice in the $(x,y)$ plane is shown, with the $z$ coordinate expressed in terms of
the instability location $z_{inst}$. Instabilities in both radial and axial directions result in
local potential minima which can retain atoms away from the center of the trap.
All quantities are expressed in arbitrary units throughout the paper.}
\label{Fig1}
\end{figure}

As clear from the expression for the magnetic field, the potential is not fully symmetric 
around the $z$-axis. Exchange in $x$ and $y$  is affected by a relative minus sign in the 
gradient term (necessary to ensure a divergenceless magnetic field), but in the small 
oscillation limit the trapping frequencies along the $x$ and $y$ axes are degenerate. 
The expressions specifying the frequencies in the three directions  are 
$\omega_x^2(0)=\omega_y^2(0)=\mu(B'^2/B_0-B''/2)/m, \omega_z^2(0)=\mu B''/m$.

The more general, spatially-dependent, frequencies can be expressed in a relatively simple form by 
introducing two characteristic lengthscales of the trap, namely $\xi=(B_0/B'')^{1/2}$ and $\eta=B_0/B'$.
Then the expressions for the effective squared frequencies become
\begin{equation}
\omega_x^2= \omega_x^2(0) \frac{1-z/[(\eta \xi^2)(1/\eta^2-1/(2\xi^2))]
+(x^2+y^2)/[8\xi^4(1/\eta^2-1/(2\xi^2))]}
{\{[1+[z^2-(x^2+y^2)/2]/(2\xi^2)]^2+[1/\eta-z/(2\xi^2)]^2x^2+[1/\eta+z/(2\xi^2)]^2y^2\}^{1/2}},
\tag{4a}
\label{4a}
\end{equation}

\begin{equation}
\omega_y^2= \omega_y^2(0) \frac{1+z/[(\eta \xi^2)(1/\eta^2-1/(2\xi^2))]
+(x^2+y^2)/[8\xi^4(1/\eta^2-1/(2\xi^2))]}
{\{[1+[z^2-(x^2+y^2)/2]/(2\xi^2)]^2+[1/\eta-z/(2\xi^2)]^2x^2+[1/\eta+z/(2\xi^2)]^2y^2\}^{1/2}},
\tag{4b}
\label{4b}
\end{equation}

\begin{equation}
\omega_z^2= \omega_z^2(0) \frac{1+[z^2-(x^2+y^2)/2]/(2\xi^2)-x^2/[2z(1/\eta-z/(2\xi^2))]+
y^2/[2z(1/\eta+z/(2\xi^2))]}
{\{[1+[z^2-(x^2+y^2)/2]/(2\xi^2)]^2+[1/\eta-z/(2\xi^2)]^2x^2+[1/\eta+z/(2\xi^2)]^2y^2\}^{1/2}}.
\tag{4c}
\label{4c}
\end{equation}

\begin{figure}[t]
\includegraphics[width=0.40\textwidth, clip=true]{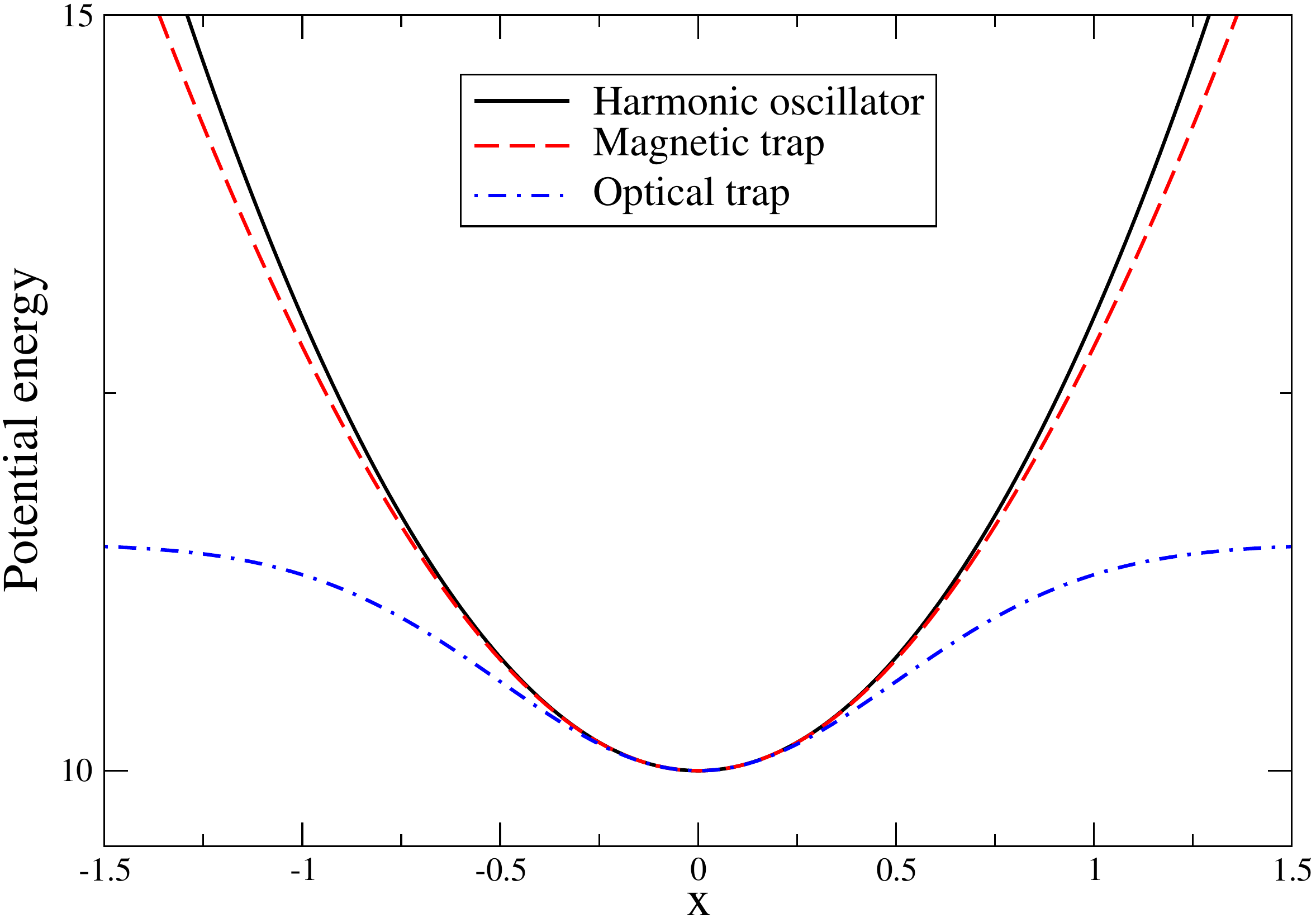}
\includegraphics[width=0.40\textwidth, clip=true]{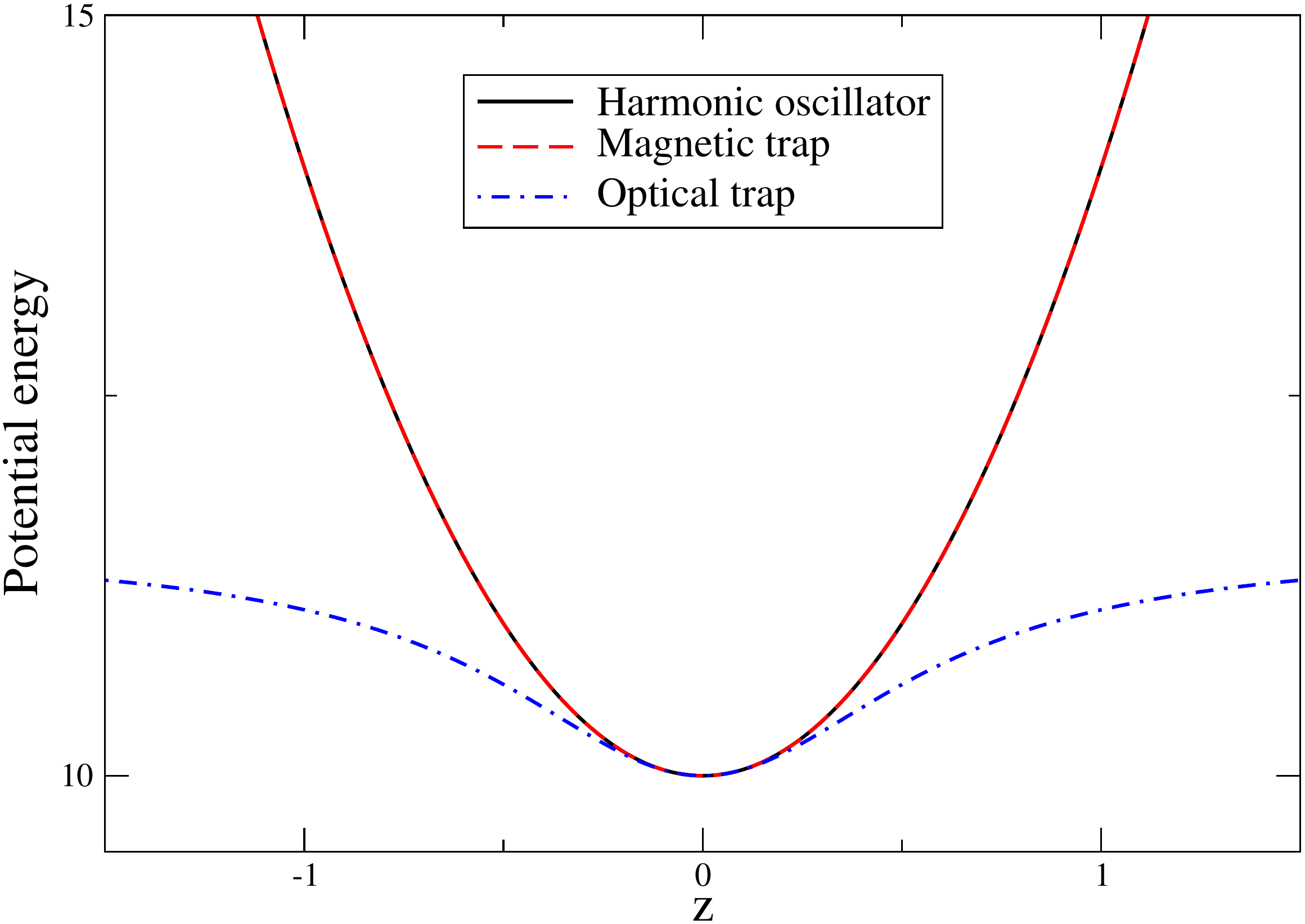}
\caption{Nonlinearities in realistic traps. The potential energy is shown versus the radial coordinate 
$x$ (left) and the axial coordinate $z$ (right) for the cases of a Ioffe-Pritchard magnetic trap 
(red, dashed curve), the ideal harmonic oscillator (black, continuous curve), and the optical dipole trap 
(blue, dot-dashed curve), all with the same small oscillation angular frequencies. 
The ideal harmonic oscillator and the optical potential have both been shifted by an amount 
$\mu B_0$ to coincide with the bottom of the magnetic trap potential. 
We have chosen $\mu=1$, $m=1$, $B_0=10$, $B'=10$, and $B''=8$.}
\label{Fig2}
\end{figure}

Given the spatial complexity of the magnetic potential, a determination of the trap depth is rather involved, 
requiring numerical evaluation. There are three input parameters characterizing the magnetic field, 
bias field $B_0$, field gradient $B'$, and field curvature $B''$, and three output parameters, trap depth, 
radial frequency, and axial frequency. As seen by inspecting Eq. (\ref{MagField}), instabilities 
in atomic trapping occur for two reasons. First, the $B_x$ or $B_y$ components change sign, which happens 
at threshold values of the $z$ coordinate $z_{inst}=\pm 2 B'/B''$, regardless of the values in the $(x,y)$ plane. 
Second, the $B_z$ component can also change sign, which occurs for all points $(x,y,z)$ such that 
$B_0+(B''/2) [z^2-\frac{1}{2}(x^2+y^2)]=0$. This can be recast as a spatial condition 
\begin{equation}
x^2+y^2=2 z^2 +4 B_0/B'' \;,
\tag{5}
\label{StabilityMT}
\end{equation}
{\it i.e.} a continuum of circles in the $(x,y)$ plane, with radii starting from 
a minimum value of $2 \sqrt{B_0/B''}$ at $z=0$, and increasing quadratically with $|z|$.
Notice that the latter instability is dynamical, as it is conditional on achieving a well-defined value of 
$z$, while the former is fixed by the parameters of the trapping potential alone, a consideration which will 
play a role in interpreting simulation outcomes later on. The resulting potential depth will be, of course, 
the minimum value between the ones resulting from these two instability conditions. An indication of the presence 
of instabilities is seen in Fig.~\ref{Fig1}, where the magnetic potential is shown in $z$ and along a specific direction 
in the $x-y$ plane. These result in local valleys in the potential which can segregate
more energetic atoms from the main clouds and affect interactions and, thus, thermalization. Note that
what is shown is the magnitude of the potential which does not entirely reflect changes in the signs of the 
magnetic force at these instabilities. In other words, the antitrapping features of the potential are not seen.   

We now shift attention to the second type of nonlinear trap we consider, a single-beam optical dipole trap (ODT) 
with potential energy
\begin{equation}
V_{\rm opt}(x,y,z)=V_0 \left[1-\frac{\exp\left(\frac{-2(x^2+y^2)}{w_0^2(1+(z/z_R)^2)}\right)}
{1+(z/z_R)^2} \right].
\tag{6}
\end{equation}
The potential is uniquely defined by the parameter $V_0$, the waist $w_0$, and the Rayleigh range 
$z_R=\pi w_0^2/\lambda_L$, where $\lambda_L$ is the laser wavelength. The potential energy amplitude $V_0$ 
is linearly related to the laser power and coincides with the trap depth, {\it i.e.} the maximum energy 
required to escape from the trap. These three parameters unambigously define the three 
trapping frequencies for small oscillations around the trap minimum. 
Due to the cylindrical symmetry around the $z$ axis two frequencies coincide, and the explicit dipole 
force acting on the atoms is expressed, in terms of its Cartesian components $(F_x, F_y, F_z)$, as

\begin{equation}
F_x=-\frac{4V_0}{w_0^2} \frac{\exp\left(\frac{-2(x^2+y^2)}{w_0^2(1+(z/z_R)^2)}\right)}
{[1+(z/z_R)^2]^2}x,
\tag{7a}
\label{forceodtx}
\end{equation}

\begin{equation}
F_y=-\frac{4V_0}{w_0^2} \frac{\exp\left(\frac{-2(x^2+y^2)}{w_0^2(1+(z/z_R)^2)}\right)}
{[1+(z/z_R)^2]^2}y,
\tag{7b}
\label{forceodty}
\end{equation}

\begin{equation}
F_z=-\frac{2 V_0}{z_R^2} \frac{\exp\left(\frac{-2(x^2+y^2)}{w_0^2(1+(z/z_R)^2)}\right)}
{[1+(z/z_R)^2]^2} \left[1- \frac{2}{w_0^2} \frac{x^2+y^2}{1+(z/z_R)^2}\right] z.
\tag{7c}
\label{forceodtz}
\end{equation} 
The small oscillation frequencies are given by
$\omega_x^2(0)=\omega_y^2(0)=4 V_0/(m w_0^2), \; \omega_z^2(0)=2 V_0/(m z_r^2)$.
The nonlinearities are factorized in Eqs. (\ref{forceodtx},\ref{forceodty},\ref{forceodtz}), and therefore the
local oscillation frequency experienced by the atoms at a generic position can be easily rewritten as the small 
oscillation frequencies multiplied by a spatially-dependent nonlinear factor. 
Figure~\ref{Fig2} illustrates the large distance deviations of the two realistic trapping potentials 
from the idealized harmonic approximation with the same small oscillation frequency. 

In general, the optical dipole trap potential deviates more from the ideal harmonic case at smaller 
distances than the magnetic trap potential. This also results in an overall change in curvature which, as 
discussed in further detail in the next section, may be detrimental to thermalization dynamics. 
Experimental groups working on all-optical cooling to achieve degenerate quantum gases are well-aware 
of this drawback of optical dipole traps~\cite{Adams}. In particular, evaporative cooling was found to become less efficient 
due to the simultaneous reduction of potential depth and curvature of the trap, the latter lowering significantly the 
elastic scattering rate necessary to repopulate the Boltzmann high-energy tail. Nevertheless, optical dipole traps are 
convenient especially as they allow for the independent use of tunable magnetic fields to control the scattering 
properties of the atoms. While this drawback has not prevented achieving Bose-Einstein condensation in the more complex 
configuration of two crossed beams~\cite{Barrett}, various techniques have been discussed to mitigate the progressive 
decrease in atomic density~\cite{Cennini,Kinoshita,Hung,Clement,Arnold}.

In our context, additional issues with ODTs occur in the peripheral regions of the trap, critical in the early stage
of evaporation and sympathetic cooling. From the expression given in Eq.~\ref{forceodtz}, it is clear that the force
in the $z-$direction can reverse sign. This occurs at thresholds given by 
\begin{equation}
x^2+y^2 = \frac{w_0^2}{2}\left[ 1+(z/z_R)^2\right] \;,
\tag{8}
\label{StabilityODT}
\end{equation}
which are, analogously to Eq. (\ref{StabilityMT}), concentric circles starting at a radius $w_0/\sqrt{2}$. 
This dynamical instability in the radial direction depends on the parameters of the trapping potential.
As discussed in the context of the numerical results, this dependence suggests a mechanism for pushing this
instability further out from the trap center which could assist thermalization.

\section{Numerical Results}

Within this framework, we have run numerical simulations using the Hamiltonian 
\begin{equation}
H_{\mathrm{tot}} =  \sum_{m=1}^{N_p} \left( \frac{P_{m}^2}{2M} +  V(Q_{m}) \right) +
\sum_{n=1}^{N_b} \left(  \frac{p_{n}^2}{2m} + V(q_n) \right) \nonumber  
+ \gamma_E \sum_{m=1}^{N_p} \sum_{n=1}^{N_b} \exp \left[-\frac{(q_n-Q_m)^2}{\lambda^2}\right],
\tag{9}
\label{hamil}
\end{equation}
where $(q_n-Q_m)^2=(x_n-X_m)^2+(y_n-Y_m)^2+(z_n-Z_m)^2$. $q_n=(x_n,y_n,z_n)$ and $Q_m=(X_m,Y_m,Z_m)$ are the
Cartesian coordinates of the particles in the two species, and $V$ is the harmonic, ODT or MT potential.
The particle motion is completely described by vectors of dimensions $6N_b$, i.e. $(q_n,p_n)$ and $6N_p$, {\it i.e.}
$(Q_m,P_m)$, respectively, where $N_b$ is the number of the coolant atoms (constituting the 'bath') and $N_p$ the number of
particles of the cooled species. The interaction potential is the one we introduced earlier~\cite{OnoSun} and is
characterized by a range $\lambda$, and a strength $\gamma_E$. Larger interaction strengths $\gamma_E$ obviously increase the 
thermalization speed. Up to a point, this occurs for larger values of $\lambda$ as well. Overall, the interaction term is 
genuinely local and nonlinear, but recovers the extended and linear case of the original Caldeira-Leggett model in the
$\lambda \rightarrow \infty$ limit, for which thermalization of single-frequency baths is strongly inhibited~\cite{Smith}. 

We start from atomic clouds at different temperatures, assuming that they are initially prepared as if at equilibrium
in a purely harmonic potential. The initial conditions for the simulations are drawn from these thermal distributions, a
starting configuration which allows for a fair comparison of all the explored traps. This is consistent with the usual
experimental protocol of precooling initial atomic clouds in a magneto-optical trap and then suddenly transferring them
into the conservative trap. As detailed in~\cite{JauOnoSun}, the inverse temperature is computed
at each time step in terms of the variance in the total energy of species, $\sigma_E^2=\langle E^2 \rangle
-\langle E \rangle^2$, using the relationship $\beta=\sqrt{D}/\sigma_E$ valid for a Boltzmann energy distribution
of a D-dimensional system. The results shown are for a single realization of the numerical experiment, which provides
information on the shot-to-shot fluctuations, influenced by the finite number of atoms,  which would be
lost on averaging over several independent runs.
Due to the relatively more complicated expressions, it is easier to choose the input parameters 
for the magnetic trap, and then derive parameters for the corresponding ideal harmonic trap and the optical 
dipole trap based on aligning the small oscillation trapping frequencies. 
We considered a wide range of parameters which can be broadly categorized in terms of the initial temperatures 
of the clouds and the relative values of the trapping frequencies in the different directions. 
At higher initial temperatures, the nonlinearities arising from the trapping potential are more relevant 
than at lower values and this serves to distinguish thermalization scenarios. The aspect ratio of the trapping 
frequencies determined by $\omega_x/\omega_z$, assuming cylindrical symmetry in keeping with experimental situations, 
changes the effective dimensionality of the particle dynamics with associated consequences for the thermalization dynamics.

\begin{figure}[t]
\includegraphics[width=0.40\textwidth, clip=true]{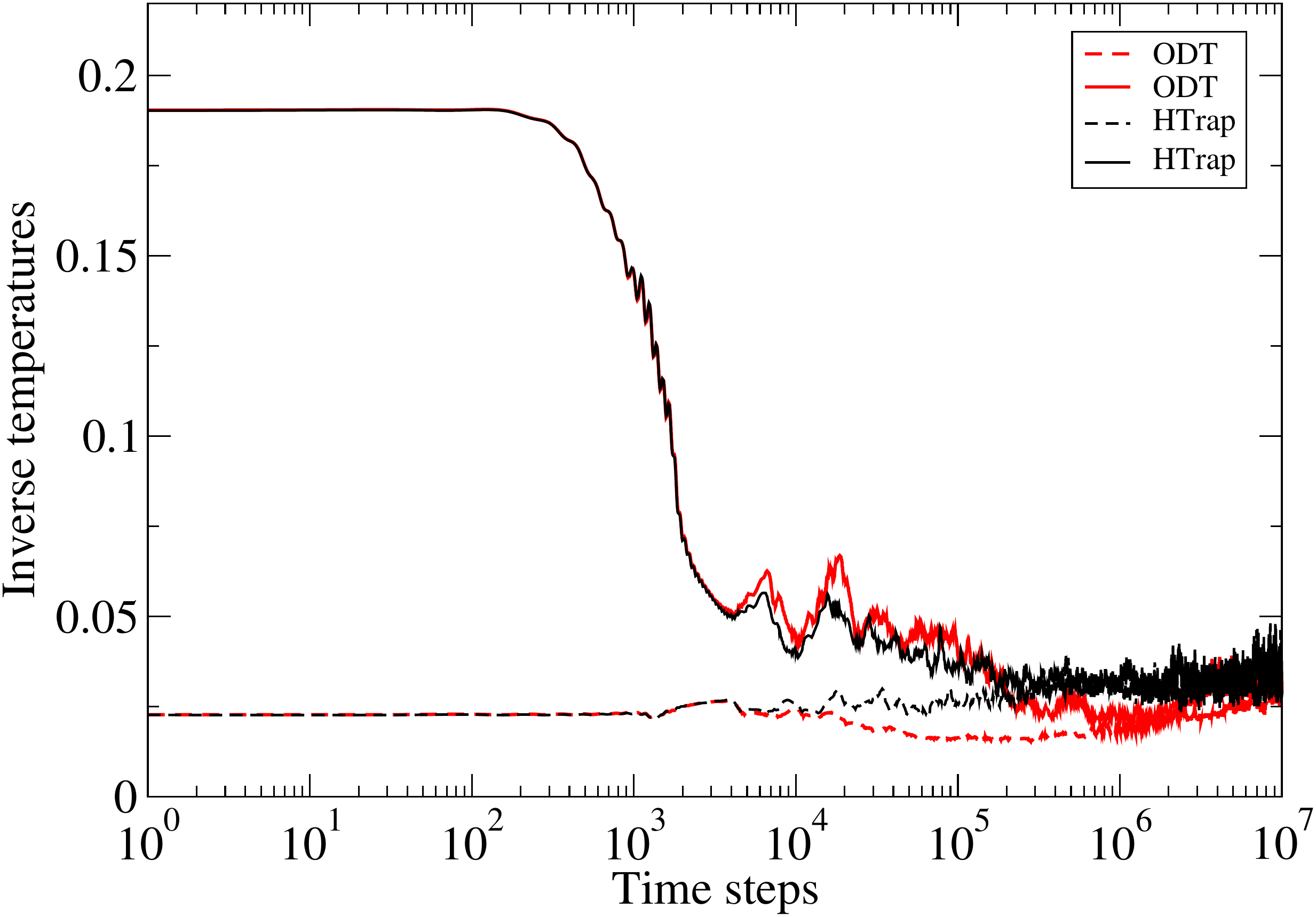} 
\includegraphics[width=0.40\textwidth, clip=true]{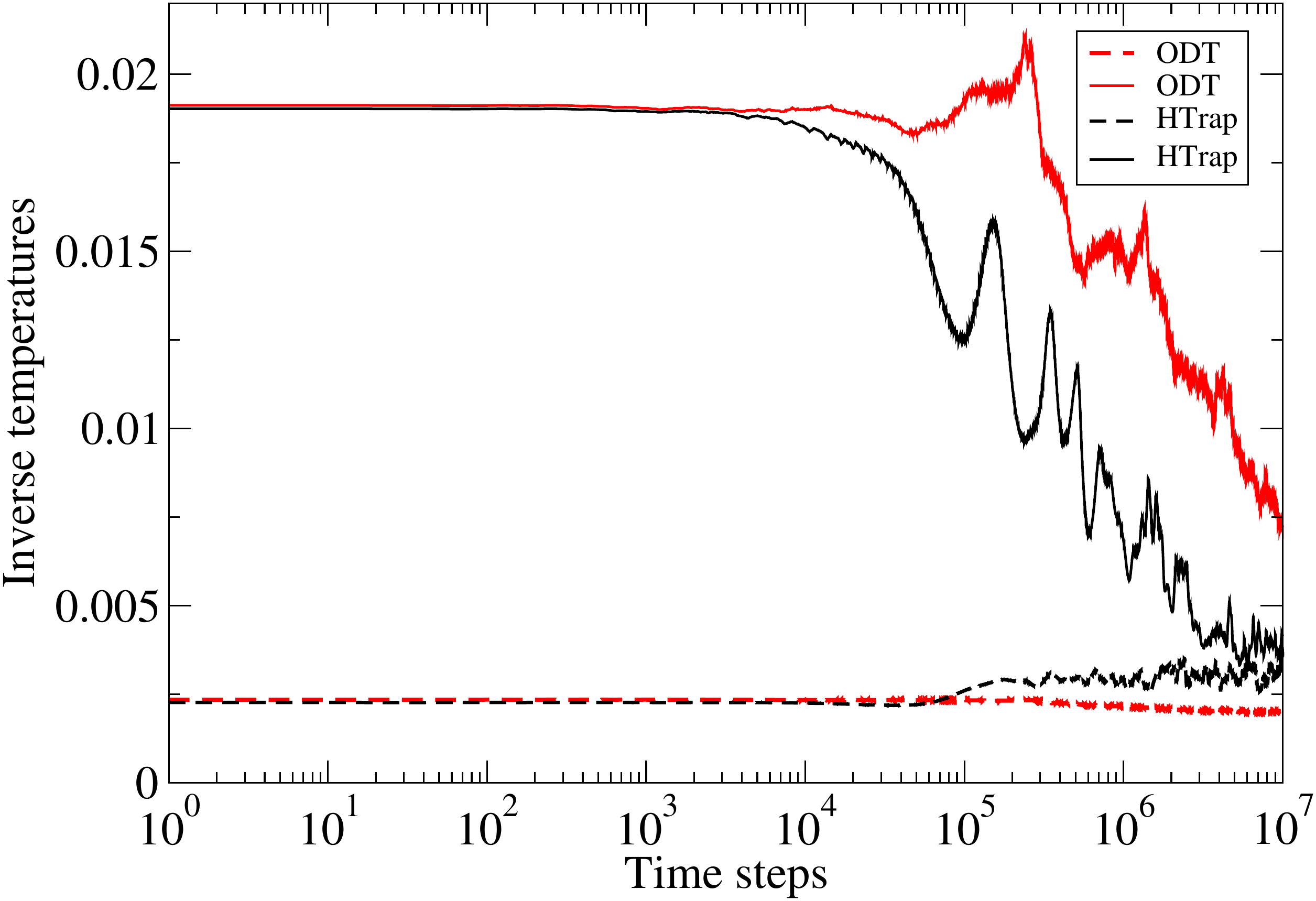}
\caption{Comparison between the thermalization dynamics for trapping potentials corresponding to an 
ideal harmonic confinement and to a realistic optical dipole trap due to a Gaussian-shaped laser beam. 
The common parameters are $N_b=N_t=100, \gamma_E=-1, \lambda=10$. In (a), the initial inverse temperatures are 
$\beta_p=0.02$ and $\beta_b=0.2$, while in (b) they are both lower by a factor of 10, {\it i.e.} $\beta_p=0.002$ and 
$\beta_b=0.02$. The parameters of the equivalent Ioffe-Pritchard trap are $B_0=20, B'=10, B''=0.01$, corresponding 
to a quasi-one dimensional trap with small oscillation frequencies $\omega_x=\omega_y=2.235$ and $\omega_z=0.1$ for 
both bath and test particles. Whereas in the low-temperature case the thermalization plots are nearly 
indistinguishable, at the higher temperatures the nonlinearity of the optical dipole trap results in 
a thermalization timescale almost one order of magnitude larger than the corresponding harmonic oscillator 
situation.} 
\label{Fig3}
\end{figure}

Since the optical dipole trap differs significantly from the ideal case as shown in Figure~\ref{Fig2}, we begin 
with a comparative analysis in the simplest case of quasi-one-dimensional dynamics, shown in Figure~\ref{Fig3}.
The two panels (a) and (b) correspond to low and high initial cloud temperatures, respectively. 
As expected, at low temperatures, there is very little to differentiate between the two traps whereas, at higher 
temperatures, the harmonic trap is clearly more efficient in terms of thermalization, with the temperatures in 
the optical dipole trap case lagging behind the ones of the ideal case.  
In the case of the harmonic potential, notice also the distinct thermalization curves for the two temperature values. 
This is due to the nonlinear and nonperturbative nature of the interaction, for the large values of 
$\gamma_E$ and $\lambda$ chosen, indicating that a scaling property with temperature is not satisfied.

\begin{figure}[t]
\includegraphics[width=0.85\textwidth, clip=true]{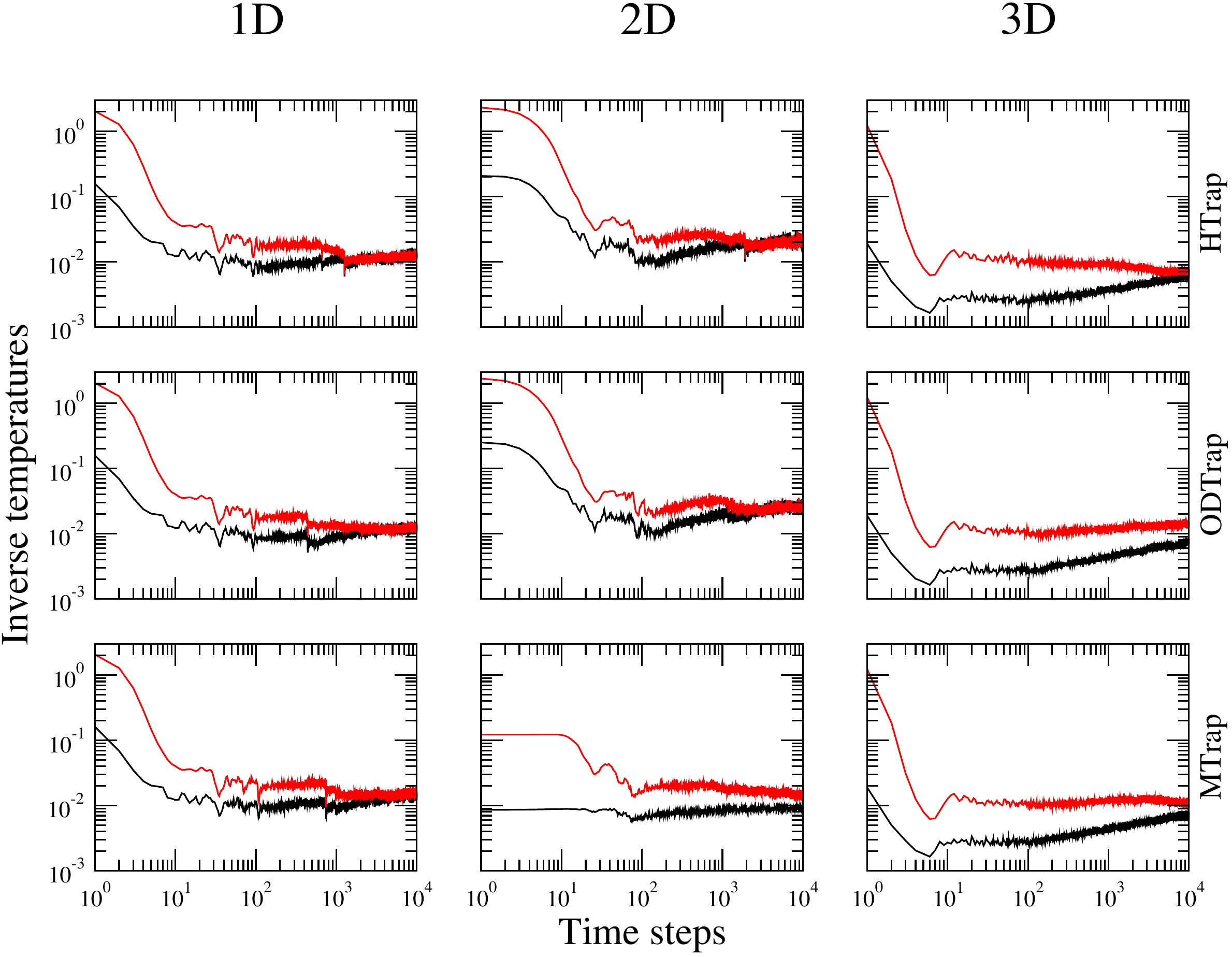}
\caption{Thermalization of mixtures in different traps and different effective dimensionality. 
The columns are relative to a comparison with a magnetic trap having $B_0=20, B'=10$ in each case but $B''$ is changed
such that the ratio of small oscillation frequencies result in different effective dimensional dynamics. The values
for the three columns are $B''=0.1, 9.8, 3.3333$ respectively where the corresponding small oscillation frequencies 
$(\omega_x,\omega_y,\omega_z)$ are $(2.22,.2.22,0.32)$, $(0.33,0.33,3.13)$ and $(1.83,1.83,1.83)$ for either type of trap. 
The mixture is balanced, $N_p=N_b=500$, with the same interatomic interaction parameters, $\gamma_E=-14, \lambda=0.5$. 
The initial inverse temperatures are $\beta_p= 0.2, \beta_b=2.0$. Note the exothermic nature of the interactions in all
the cases shown, and the abrupt drop in inverse temperatures for the 2D magnetic trap case, as discussed in the text. }
\label{Fig4}
\end{figure}

\begin{figure}[bt]
\includegraphics[width=0.85\textwidth, clip=true]{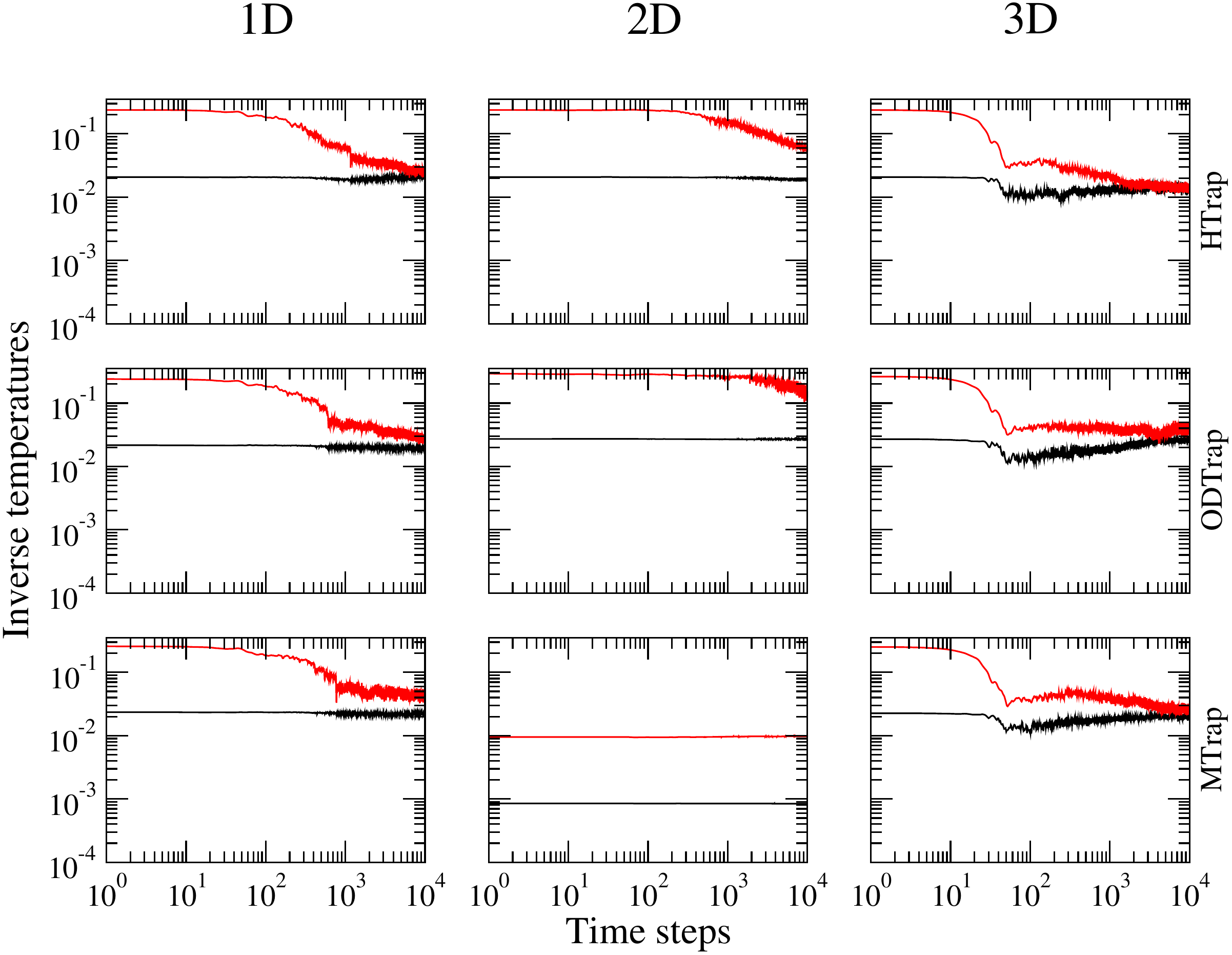}
\caption{Same conditions as in Figure~\ref{Fig4} except that the initial conditions correspond to
hotter temperatures. Here the starting inverse temperatures are $\beta_p= 0.02, \beta_b=0.2$. Unlike
at lower temperatures, the interaction is no longer manifestly exothermic, apart from the 2D magnetic 
trap case for which however thermalization is absent.}
\label{Fig5}
\end{figure}

Next, we include the effect of dimensionality and the case of the magnetic trap.
Figures~\ref{Fig4} and ~\ref{Fig5} show results contrasting all three types of traps at low and high initial
temperatures respectively, but for distinct frequency ratios. As seen from the figure captions, the triplet  
of magnetic field parameters $(B_0, B', B'')$ was chosen to stiffen the traps in one or more directions. 
The effective dimensionality of the dynamics is indicated at the top of each column. 
The low temperature results are clearly distinct from those at higher initial temperatures. 

At lower initial temperatures (Fig.~\ref{Fig4}), the fact that for all effective dimensionalities the final convergent
temperature is hotter than either of the initial ones is a clear indicator of the dominance of the interaction 
term, analogous to exothermic situations discussed earlier~\cite{OnoSun,JauOnoSun}. 
For effective 1D dynamics, there is little to differentiate between the three trap configurations.
In the 2D case, the MT shows signs of delayed convergence while the ODT closely follows the pure harmonic dynamics.
Furthermore, this case shows an initial drop in inverse temperatures of both species which is not seen in any of
the other cases. In the 3D situation, there are differences at late time while the early time dynamics are virtually identical.
In particular, the ODT seems slightly less efficient with regard to thermalization, presumably due to the shape of the potential
which differs significantly from the harmonic trap even at moderate distances from the center, as discussed in the context
of Fig.~\ref{Fig2}. 
 
The origins of the anomalous behavior observed in the MT 2D case arises from the instabilities, in $z$ and radial 
directions, alluded to earlier. For the stated magnetic field conditions, the threshold values for both instabilities 
is the least (closest to the origin) for the 2D situation. In fact, even at these low temperatures, substantial 
fractions of the initial conditions for both species lie beyond the threshold for the radial instability. 
These fractional values are $0.84$ and $0.34$ for target and bath species respectively. By contrast, the 
corresponding values in 1D and 3D are negligibly small for both instabilities. Also, the fractions 
beyond the threshold of the axial instability are $0.012$ and essentially zero for the 2D case. 
The radial instability appears to impact early time behavior while not dramatically altering 
thermalization at longer times.

For the higher initial temperatures shown in Fig.~\ref{Fig5}, even in the effective 1D situation, there are 
clear differences between the three types of traps with the harmonic one still being the most efficient. 
The converged temperature, in most of the cases shown, is trending towards a value between the two initial ones, 
or at least does not appear as exothermic as in the lower temperature case. For the harmonic trap, this is a consequence 
of the relative importance of the interaction energy as compared with the energies of the individual clouds, where 
the latter is dominant at these higher temperatures. Moreover, the spatially extended initial conditions at higher 
temperatures better explore the nonlinear features of the MT and ODT potentials.  
Two additional features are also worth noting. 

\begin{figure}[tb]
\includegraphics[width=0.65\textwidth, clip=true]{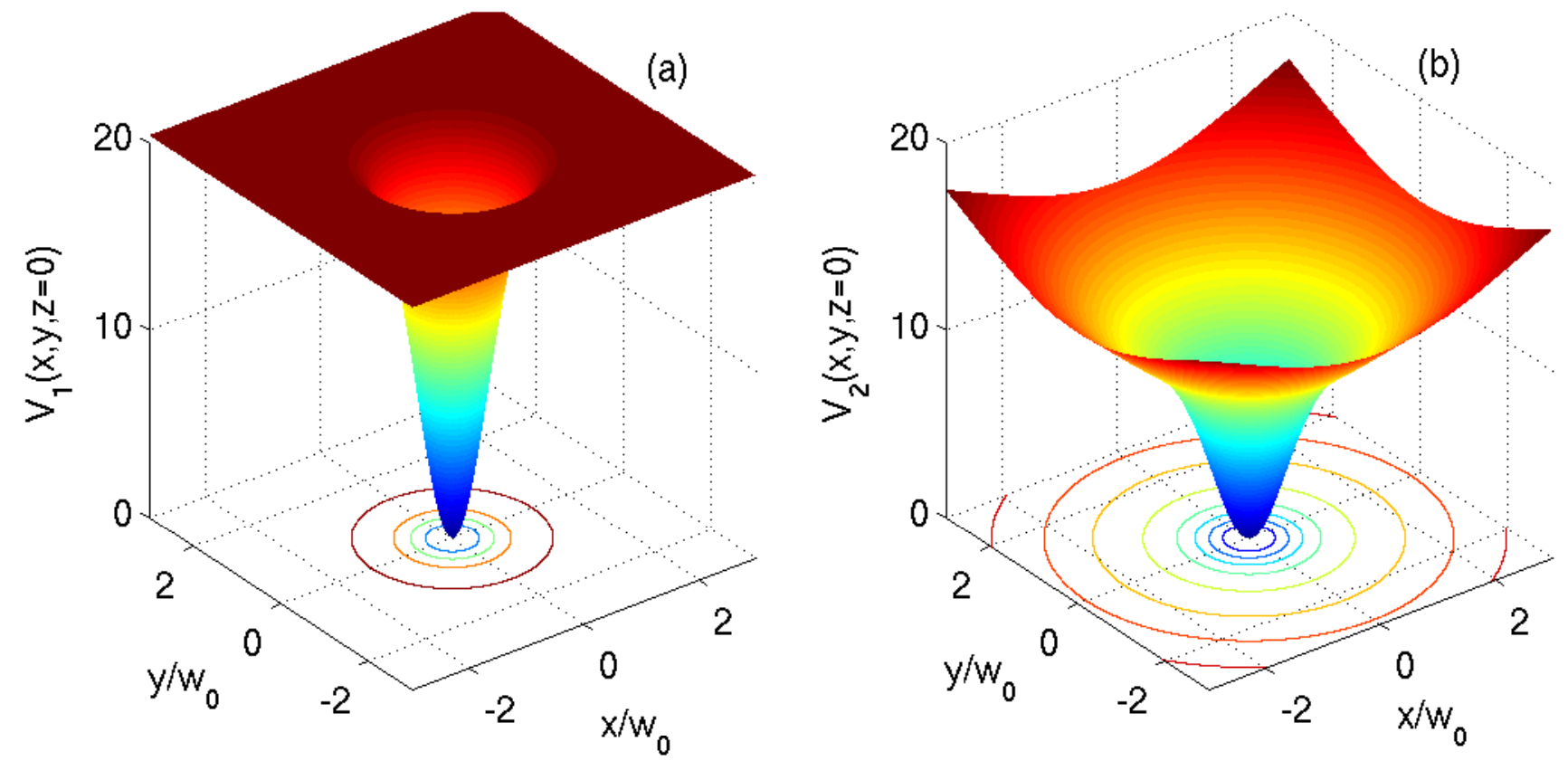}
\includegraphics[width=0.65\textwidth, clip=true]{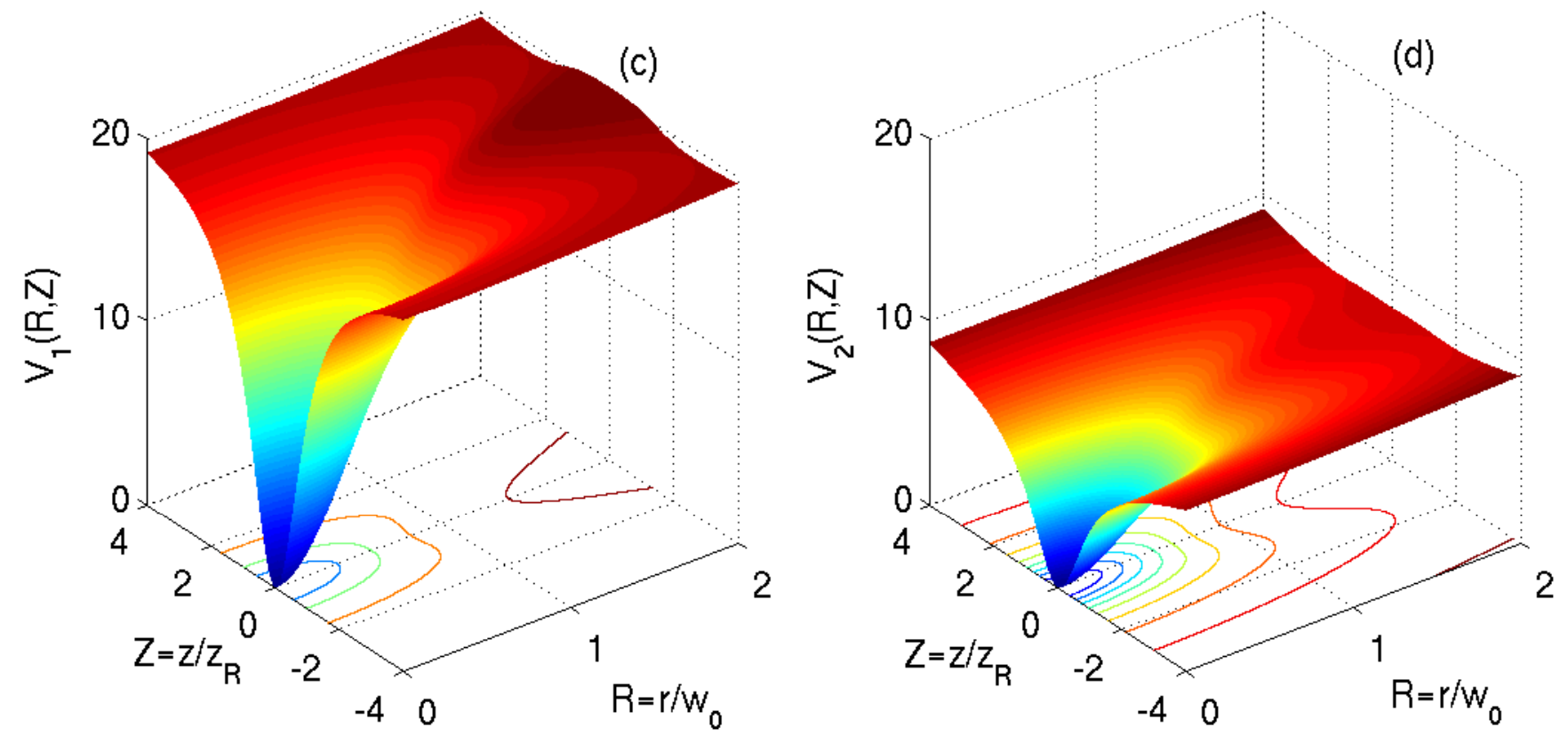}
\caption{Contrasting the usual ODT (a) with a more general double optical dipole trap (DODT)
(b) for $z=0$, as discussed in the text, with $c_0=0.4$ and $c_1=0.6$. $(x,y)$ are expressed in
terms of the beam waist. The presence of a second laser beam increases the retaining capability in the outer region 
of confinement in which the single optical dipole trap has a nearly flat potential. Panels (c) and (d) show the same 
two cases using the cylindrical symmetry, with $r$ relative to the waist and $z$ to the Rayleigh range $z_R$.
The radial instability given by the condition Eq. (\ref{StabilityODT}) is clearly visible in (c) while the fact 
that the secondary beam has a waist which is five times the one of the primary beam pushes the instability further out in (d). 
Both waist and Rayleigh range of the auxiliary beam preserve the aspect ratio of the single trap.}
\label{Fig6}
\end{figure}

First, the 3D cases thermalize better than the corresponding 1D instances. This is somewhat surprising as the 
expectation is that, in higher dimensions, the ability to avoid interactions should be enhanced. 
Our earlier results required higher values of $\gamma_E$ for higher dimensions to achieve comparable 
thermalization timescales~\cite{JauOnoSun}, so we anticipated the 1D case to be even more efficient. 
A possible interpretation of this outcome is the role played by trapping regions in which the particle 
motion is chaotic. Previous work has explored the role of chaos for evaporative cooling in magnetic traps,
concluding that the effective dimensionality of the process changes from being fully three-dimensional
to one-dimensional as one moves from chaotic to integrable trajectories ~\cite{Surkov1,Surkov2,Pinkse,Harms}.
Further, studies of magnetic traps~\cite{Salas} and crossed optical dipole traps~\cite{Gonzalez} confirm
the presence of chaotic trajectories for the most energetic particles. 

Second, the 2D situation is clearly the anomaly with respect to the 1D and 3D cases and the MT 2D case is 
the most distinct as there are no signs of thermalization whatsoever. Contrasted with the low temperature
case, not only are higher fractions of initial conditions above the radial threshold (now $0.97$ and $0.84$
for particles and bath respectively) but a significant fraction of the particles ($0.52$) are also above the
threshold for the axial instability. By comparison, these fractions in 1D are negligible, while the 3D values
are no larger than $0.28$. As indicated earlier, we infer from this that large fractions beyond the radial
instability affect the early time dynamics, with heating leading to an abrupt drop in inverse temperatures
immediately after the simulation starts. To reiterate, this is a consequence of the hotter atoms, associated
 with the higher initial temperatures, which are transferred from the precooling stage (typically from a magneto-optical
 trap) in which, as discussed above, harmonic trapping is assumed. Unlike the earlier situations, this effect is
 further compounded with the onset of the axial instability. We have contrasted trajectories, starting from initial
 conditions on either side of the instabilities, to gauge the impact of the two kinds of instabilities discussed.
 Similar runs were also used to extract Lyapunov exponents which clearly show the role of chaos in the dynamics above
the threshold for instability.

The presence of threshold values for antitrapping regions, as in the MT case, or the very presence of trajectories 
quite far apart from the center of the trapping potential in general,  strongly affect the thermalization efficiency. 
It is possible to mitigate this effect by adding an auxiliary potential having minimal effect near the center, while
forcing atoms in the peripheral regions towards the center of the trap. This addition would also assist in the case
of ODT where the softening of the potential, as seen in Fig. \ref{Fig2}, can lead to shelving of fractions of the populations. 

\begin{figure}[t]
\includegraphics[width=0.65\textwidth, clip=true]{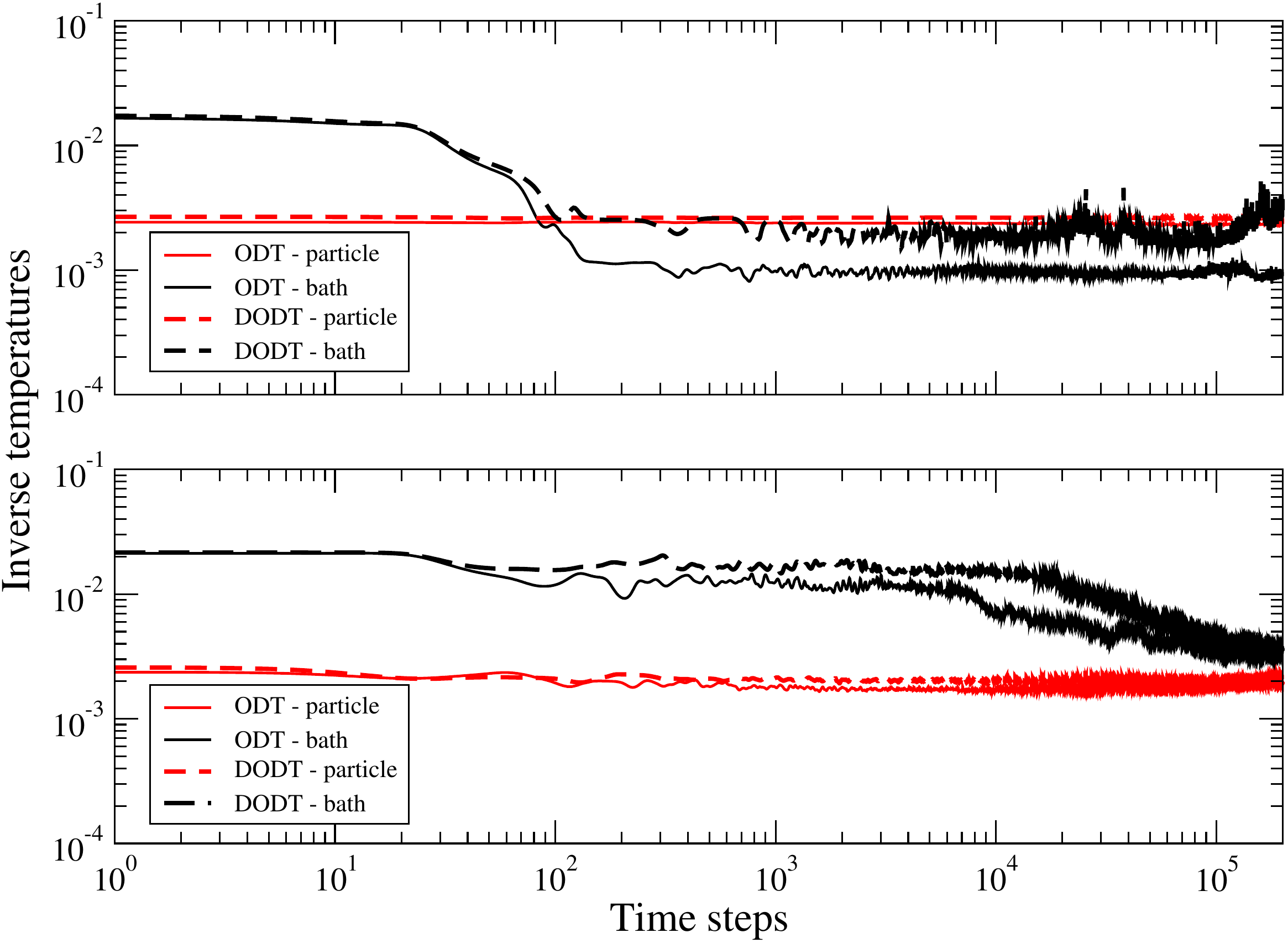}
\caption{Thermalization dynamics for optical dipole traps with and without an auxiliary laser beam with large 
waist, assisting the recapture of more energetic atoms. The top plot is relative to an unbalanced 
mixture in which $N_b=10$, $N_p=390$, the bottom plot for a balanced mixture ($N_b=N_p=200$). 
Other parameters are common and equal to $\gamma_E=+14$ (corresponding to repulsive interspecies 
interactions enhancing the shelving effect), $\lambda=10$, $\beta_p=0.002$, $\beta_b=0.02$, $c_0=0.4$, $c_1=0.6$. 
The small oscillation frequencies for both ODT and DODT are $(\omega_x, \omega_y, \omega_z)=(6.95,6.95,1.83)$.}
\label{Fig7}
\end{figure}

A secondary potential can be realized in the case of an optical dipole trap by splitting the original laser 
beam into two components, and sending the two beams with two different focal lengths on the center of the 
optical dipole trap. The laser beam corresponding to minimal beam waist is aimed at maintaining a strong 
confinement at the center, while the one with larger waist gives rise to a stiffer potential on the periphery. 
We consider two laser beams with waists $w_{0}$ and $w_{1}$ and wavelengths 
$\lambda_{L0}$ and $\lambda_{L1}$. The Rayleigh ranges are then $z_{0}=\pi w_{0}^2/\lambda_{L0}$ and 
$z_{1}=\pi w_{1}^2/\lambda_{L1}$. We start with the simplifying hypothesis that the laser is common 
to the two beams, so we have $\lambda_{L0}=\lambda_{L1}$. The generic optical dipole potential is
\begin{equation}
V_{\rm opt}(x,y,z)=V_0
\left[
c_0 + c_1 -
c_0 \frac{\exp\left(\frac{-2(x^2+y^2)}{w_{0}^2(1+(z/z_{0})^2)}\right)}{1+(z/z_{0})^2}-
c_1  \frac{\exp\left(\frac{-2(x^2+y^2)}{w_{1}^2(1+(z/z_{1})^2)}\right)}{1+(z/z_{1})^2}
\right].
\tag{10}
\end{equation}
The arbitrary offset of the potential is chosen in such a way that $V(0,0,0)=0$ and
its asymptotic value at infinite distance from the origin is  $V(|\infty|)/V_0=c_0+c_1$.
In general, the expressions for the forces along the three directions can be readily obtained. 
These expressions, once linearized near the origin, give for the three Cartesian components of the force

\begin{equation}
F_x=-4V_0  \left(\frac{c_0}{w_0^2}+\frac{c_1}{w_1^2} \right)x,
\tag{11a}
\end{equation}

\begin{equation}
F_y=-4V_0  \left(\frac{c_0}{w_0^2}+\frac{c_1}{w_1^2} \right)y,
\tag{11b}
\end{equation}

\begin{equation}
F_z=-2 V_0 \left(\frac{c_0}{z_0^2}+\frac{c_1}{z_1^2} \right)z.
\tag{11c}
\end{equation}

Since we are interested in comparing the behaviour of the traps far from the center, we have 
chosen the parameters $c_0$ and $c_1$ and the laser waists such that the small oscillation frequencies 
for the ODT and double optical dipole trap (DODT) are the same. This more generalized potential is contrasted 
with the usual ODT in Fig.~\ref{Fig6}. Our choice implies more laser power in the DODT and  
hence a larger energy depth which is reflected in the continued growth of tails of the potential. 
Figure~\ref{Fig7} contrasts the thermalization curves for the ODT and DODT cases in two situations;
where the two species have equal numbers of particles and the other where there are substantionally fewer
cooler bath particles. The initial temperatures of both clouds are chosen to be higher 
than the hottest cases analyzed so far in Fig.~\ref{Fig5} to enhance the shelving effect and show 
the impact of the additional ODT.
In the asymmetric particle case, the overall effect of the DODT is amplified with respect to the equal number mixture.
We believe this is due to the fact that, for a single ODT, the small number of the cold species are quickly ejected to the
flat peripheral regions of the potential due to their repulsive interactions with the considerably larger number of hot particles.
Once this occurs, they have only very limited encounters with the other species, the condition needed for thermalization. 
The DODT provides the mechanism to reinject them towards the trap center resulting in renewed interactions. 
Also this would lead one to expect variability in the results with initial configuration of the colder particles. When the particle numbers
are balanced, this mechanism is far less important and the thermalization timescales are comparable in the two cases. 
It would be interesting to explore the situation more relevant for cooling atoms, with a large cold reservoir of atoms trying 
to drive to lower temperatures a smaller number of hot atoms, though the particle numbers required to show this are beyond our
current computational capabilities.

Analogous considerations can be applied to the case of a MT. The addition of a 
shallow, attractive optical potential should help return atoms more quickly to the minimum of the potential.
However, the expression of the magnetic field in Eq.(\ref{MagField}) contains only the leading term in a 
multipole expansion~\cite{Bergeman} which is not enough to capture the long-distance 
behavior of the trapped atoms. Detailed {\it ab initio} evaluations of the magnetic field are required to confirm 
that the lack of thermalization intrinsic to the high-temperature situation for the 2D case in Figure~\ref{Fig5} is
mitigated by the presence of a monotonically attractive optical potential term.

\section{Conclusions}

In conclusion, we have discussed the dynamics of thermalization of atomic mixtures for realistic atomic traps. 
The interplay between nonlinearities arising from the interatomic potential term, necessary to enforce
thermalization, and the ones arising from nonlinear trapping, present in any realistic setup, has been emphasized. 
Comparison with the ideal harmonic trapping allows us to disentangle
features arising from the nonlinearity of the realistic potentials.
Thermalization is achieved in a large number of cases we have analyzed, with the notable exception of the 
magnetic trapping configuration in 2D, where instabilities in the potential result in atoms moving away from
regions of maximal interaction. This feature is clearly more detrimental at higher initial temperatures. 
We have also evidence for faster thermalization in 3D with respect to the corresponding 1D case, despite the inhibition of 
head-on collisions, an effect that may be influenced by dynamical chaos. 
The issue of whether the presence of chaotic behavior in nonlinear traps is advantageous to increasing the 
thermalization speed requires further exploration. However, our results do suggest that nonlinearities in 
atom traps may be turned into assets, as discussed also in Ref~\cite{Makhalov} with regard to precision 
measurements of the trap parameters on exploiting nonlinearities of the ODT.
Simulations like the ones shown, appropriate for each specific experimental setup, will help
identify parameters for which this effect can contribute significantly to the earlier stage 
of atomic cooling, especially during transfer of atoms from a magneto-optical trap into a conservative trap.

\end{document}